\documentclass[a4paper,11pt]{article}
\usepackage{aaskaiid}
\usepackage{subcaption} 
\usepackage{amsmath}
\usepackage{orcidlink}

\title{Low-Frequency VLBI Network Using SKA-LOW}
\ShortTitle{SKA-LOW VLBI}

\author[1,2,4]{Hideyuki Kobayashi\orcidlink{0000-0001-8066-1631}}
\ShortName{H. Kobayashi et al.} 
\author[2]{Tomoaki Oyama\orcidlink{0000-0003-4046-2923}}
\author[3]{Kazuma Fujita}
\author[2]{Syunsaku Suzuki}
\author[4]{Shintaro Yoshiura\orcidlink{0000-0003-0581-5973}}
\author[5]{Hiroaki Misawa\orcidlink{0009-0002-1091-7907}}
\author[5]{Fuminori Tsuchiya\orcidlink{0000-0003-3386-6794}}
\author[6]{Kazuhiro Takefuji\orcidlink{0000-0002-6641-755X}}
\author[7]{Kenta Fujisawa\orcidlink{0009-0008-1070-4411}}
\author[3]{Kohtaro Niinuma\orcidlink{0000-0002-8169-3579}}
\author[8]{Keitaro Takahashi\orcidlink{0000-0002-3024-5769}}
\author[9]{Yashwant Gupta\orcidlink{0000-0001-5765-0619}}
\author[9]{Bhal Chandra Joshi\orcidlink{0000-0001-5765-0619}}
\author[9]{Vishwesh Marthi\orcidlink{}\orcidlink{0000-0002-4629-314X}}
\author[10]{M. A. Krishnakumar\orcidlink{}\orcidlink{0000-0003-4528-2745}}
\author[11]{Zhiqiang Shen\orcidlink{}\orcidlink{0000-0003-0804-2478}}
\author[11]{Wu Jiang\orcidlink{}\orcidlink{0000-0001-5153-5871}}
\author[12]{Yihua Yan\orcidlink{}\orcidlink{0000-0002-7106-6029}}
\author[12]{Wei Wang}
\author[12]{Linjie Chen\orcidlink{0000-0002-3618-3430}}

\affiliation[1]{Center for Radio Astronomy and Engineering, National Astronomical Research Institute of Thailand,260 Moo 4, Don Kaeo Subdistrict, Mae Rim District, Chiang Mai 50180, Thailand}
\emailAdd{hd.kobayashi@outlook.com}
\affiliation[2]{ Mizusawa VLBI Observatory, National Astronomical Observatory of Japan, 2-12 Hoshiga-oka, Mizusawa Oshu-shi, Iwate 023-0861, Japan}
\affiliation[3]{Graduate School of Sciences and Technology for Innovation, Yamaguchi University, Yoshida 1677-1, Yamaguchi 753-8512,Japan}
\affiliation[4]{Mizusawa VLBI Observatory, National Astronomical Observatory of Japan, 2-21 Osawa, Mitaka, Tokyo 181-8588, Japan}
\affiliation[5]{Planetary Plasma \& Atmospheric Reseaerch Center, Graduate School of Science, Tohoku University, 6-3 Aramaki Aza Aoba, Sendai 980-8578, Japan}
\affiliation[6]{Tracking and Communications Center, Japan Aerospace Exploration Agency, 1831-6. Ohmagari Kami-Odagiri Saku-shi, Nagano 884-0306, Japan}
\affiliation[7]{Research Institute for Time Studies, Yamaguchi University, Yoshida 1677-1, Yamaguchi 753-8512,Japan}
\affiliation[8]{Faculty of Advanced Science and Technology, Kumamoto University, 2-39-1 Kurokami, Kumamoto 860-8555, Japan }
\affiliation[9]{National Centre for Radio Astrophysics, Tata Institute of Fundamental Research, Savitribai Phule Pune University Campus, Pune 411 007, India}
\affiliation[10]{Radio Astronomy Centre, Tata Institute of Fundamental Research, Udhagamandalam - 643001,Tamil Nadu, India}
\affiliation[11]{Shanghai Astronomical Observatory, Chinese Academy of Sciences, 80 Nandan Road, Shanghai 200030, China}
\affiliation[12]{National Space Science Center,Chinese Academy of Sciences, No.1 Nanertiao Zhongguancun Haidian District, Beijing 100190, China}

\abstract{We propose the development of a low-frequency Very Long Baseline Interferometry (VLBI) network operating in the 100–350 MHz range, incorporating the Square Kilometre Array Low (SKA-LOW). SKA-LOW is expected to achieve exceptionally high sensitivity within this frequency band. By integrating SKA-LOW with other high-sensitivity radio telescopes located across the Asia-Pacific region, the proposed network is anticipated to deliver up to two orders of magnitude improvement in sensitivity compared to the existing VLBA.
While several scientific themes utilizing low-frequency VLBI have already been proposed, we specifically advocate for astrometric studies employing existing VLBI stations to demonstrate the feasibility and scientific potential of this frequency regime. Furthermore, the combination of SKA-LOW with additional radio telescopes will enable high-fidelity imaging observations, significantly enhancing the quality and scope of low-frequency VLBI science.}

\begin{document}
\maketitle

\section{Scientific Significance and Objectives}
Modern astronomy has expanded its horizons by observing the universe in wavelengths beyond visible light—such as radio, infrared, X-rays, and gamma rays—since the 20th century. However, with the diversification of observational instruments in recent years, these frontier fields are becoming increasingly saturated. As a renewed frontier in radio astronomy, low-frequency observations (50–350 MHz) are now being explored.
VLBI observations in this frequency range have been relatively rare compared to other bands and have not received much attention. However, with the advent of new facilities such as SKA-LOW and uGMRT, new research topics are being considered. We are currently investigating the following key areas:

A) Foreground Removal for Cosmic Reionization Observations
Detecting HI emissions from the epoch of reionization (\cite{pritchard2012cosmology};  \cite{shimabukuro2023exploring}) requires highly accurate removal of foreground emissions, which constitute a major source of systematic error (\cite{barry2022mwa}). To achieve this, precise determination of the positions and flux densities of bright point sources is essential. While it is possible to estimate these using catalogs observed at different frequencies (e.g., L-band), observations in the actual SKA-LOW frequency range are ideal for reducing uncertainties. Therefore, it is necessary to construct a VLBI network including SKA-LOW and to create a source catalog. The required precision includes positional accuracy on the order of milliarcseconds and flux density accuracy within 1\% (\cite{datta2010bright}), which can only be achieved through VLBI observations.

B) Primordial Black Holes from Early-Universe Gravitational Waves
Evidence for gravitational waves from primordial sources has been obtained through Pulsar Timing Arrays (PTAs), and detection is considered imminent (\cite{agazie2023nanograv} ; \cite{EPTA_DR2_III_2023}; \cite{Reardon2023_PPTA_GWB}; \cite{Xu2023_CPTA_DR1}; \cite{Agazie2024_PTA_comparison}; \cite{Miles2025_MPTA_DR45}). In 2023, Global PTA reported an evidence of primodial gravitational wave (\cite{taylor2023methods}), which is believed to be caused by mergers of black holes formed in the early universe. Identifying the sources of these gravitational waves is crucial, and the next step is to determine their directions based on timing variations in pulsars caused by the background gravitational waves. For this, parallax measurements with 1 pc accuracy for pulsars within 100 pc are required (\cite{kato2023localization}). This level of precision can only be achieved through VLBI observations, making it a promising research area in the SKA era.
Conventional pulsar astrometry observations have primarily been conducted in the L-band. However, for millisecond pulsars (MSPs), the flux density peaks around the 320 MHz band, which is typically an order of magnitude stronger than in the L-band (\cite{Aggarwal2022MSPspectra}). Consequently, the realization of SKA-Low is highly likely to enable us to achieve the required astrometric precision even within the 320 MHz band.

Other science cases for VLBI observations using SKA-LOW include pulsar studies and investigations of interstellar plasma scintillation, among others, as discussed in the SKA Science Book (\cite{AASKA14}).

\section{Possible low frequency VLBI network}

As shown in Figure 1, we propose a VLBI network consisting of six stations currently operating in Asia: Iitate, GMRT, Ooty, FAST, and IPS array, with SKA LOW planned for the future. Furthermore, by including Narrabri, Parkes, Ceduna, and Hobert of the LAMBDA network, which is currently under development, the performance of the network will be further enhanced.

\begin{figure}[h]
    \centering
	\includegraphics[width=0.8\columnwidth]{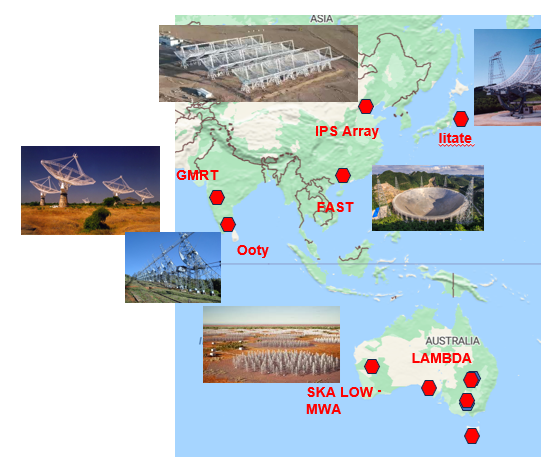}
    \caption{Array configuration of low frequency VLBI with SKA-LOW, GMRT, Ooty, IPT array. FAST, Iitate, and LAMBDA (Narrbri, Parkes, Ceduna, and Hobart)}
    \label{figure 1}
\end{figure}

Table 1  summarizes the baseline lengths between the stations. The longest baseline exceeds 9,000 km, while the shortest is approximately 300 km. The corresponding fringe spacings at a frequency of 320 MHz are about 21 and 600 milli-arc-seconds (mas), respectively. The longest baseline is 9,115 km between GMRT and Hobart, while the shortest baseline is 322 km between Narrabri and Parkes of LAMBDA. The array spans approximately 8,000 km in both the east–west and north–south directions.

\begin{table}[htpb]
	\centering
	\caption{Baseline length between stations (km)}
	\label{tab:statio_distance}
	\begin{tabular}{lcccccccccr} 
		\hline
		    & SKA-LOW & GMRT & Ooty & Iitate & FAST & IPS & Narrabri & Parkes & Ceduna & Hobart\\
		\hline
		SKA-LOW & - & 6540 & 5831 & 7155 & 5712 & 7227 & 3196 & 3083 & 1747 & 3260 \\
		GMRT & 6540 & - & 902 & 6399 & 2402 & 4528 & 8833 & 8839 & 7895 & 9115 \\
 		Ooty & 5831 & 902 & - & 6624 & 3502 & 4928 & 8312 & 8297 & 7275 & 8532 \\
   		Iitate & 7155 & 6399 & 6624 & - & 3407 & 2226 & 7164 & 7394 & 7286 & 8241 \\
		FAST & 5712 & 3404 & 3502 & 3407 & - & 1990 & 7239 & 7360 & 6648 & 8007 \\
		IPS & 7227 & 4528 & 4928 & 2226 & 1990 & - & 8121 & 8288 & 7849 & 8996 \\
        Narrabri & 3196 & 8833 & 8312 & 7164 & 7239 & 8121 & - & 322 & 1508 & 1396 \\
        Parkes & 3083 & 8839 & 8297 & 7394 & 7360 & 8288 & 322 & - & 1361 & 1089 \\
        Ceduna & 1747 & 7895 & 7275 & 7286 & 6648 & 7849 & 1508 & 1361 & - & 1703 \\
        Hobart & 3260 & 9115 & 8532 & 8241 & 8007 & 8996 & 1396 & 1089 & 1703 & - \\
		\hline
	\end{tabular}
\end{table}

Table 2 shows the System Equivalent Flux Density (SEFD) of each station and the mutual SEFDs. The SEFD of each station was estimated from reference papers for individual stations described in Section 3. Compared to conventional VLBI observations in this frequency band, such as those with the VLBA, this network offers a sensitivity improvement of 10 to 100 times.

\begin{table}[htbp]
	\centering
	\caption{Auto and mutual SEFD at 320MHz (Jy)}
	\label{tab:station_sefd}
	\begin{tabular}{lcccccccr} 
		\hline
		    & SKA-LOW & GMRT & Ooty & Iitate & FAST & IPS & LAMBDA &VLBA\\
		\hline
		SKA-LOW & 5 & 7 & 16 & 52 & 5 & 12 & 112& 117\\
		GMRT & 7 & 11 & 23 & 76 & 8 & 18 & 166 & 174\\
 		Ooty & 16 & 23 & 50 & 163 & 17 & 37 & 354 & 370 \\
   		Iitate & 52 & 76 & 163 & 531 & 55 & 122 & 1152 & 1207\\
		FAST & 5 & 8 & 17 & 55 & 6 & 13 & 118 & 124\\
		IPS & 12 & 18 & 37 & 122 & 13 & 28  & 265 & 277\\
        LAMBDA & 112 & 166 & 354 & 1152 & 118 & 265 & 2500 & 2618\\
        VLBA & 117 & 174 & 370 & 1207 & 124 & 277 & 2618 & 2742\\
		\hline
	\end{tabular}
\end{table}

LOFAR is a prominent northern-hemisphere array capable of achieving high sensitivity and angular resolution in this frequency band. Operating in the 15–240 MHz band with a maximum baseline length of approximately 2000 km, the upgraded LOFAR2.0 (\cite{hessels2023lofar2}) is projected to achieve a sensitivity of 30 $\mu$Jy/beam and an angular resolution of 0.2 arcseconds in the High Band Array with an 8-hour integration. The proposed VLBI array centered around SKA-Low will establish a complementary relationship with LOFAR, filling the reciprocal observational gaps

Figure 2 shows the UV coverage of an array consisting of Iidate, GMRT, and Ooty—stations already capable of observation—along with SKA-LOW. The maximum baseline length reaches \(8x10^8\) wavelengths, enabling a spatial resolution of approximately 25 mas at 320 MHz. As suggested by the UV coverage, the dynamic range for imaging observations is limited, but the antenna configuration is considered well-suited for astrometry observations.

\begin{figure}[htbp]
  \centering
  \begin{subfigure}[b]{0.3\textwidth}
    \centering
    \includegraphics[width=\textwidth]{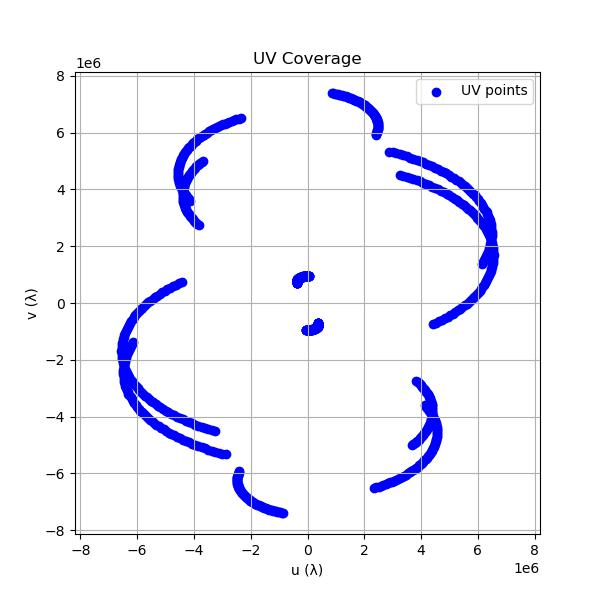}
    \caption{Dec. $+30$ deg.}
    \label{fig:fig1}
  \end{subfigure}
  \begin{subfigure}[b]{0.3\textwidth}
    \centering
    \includegraphics[width=\textwidth]{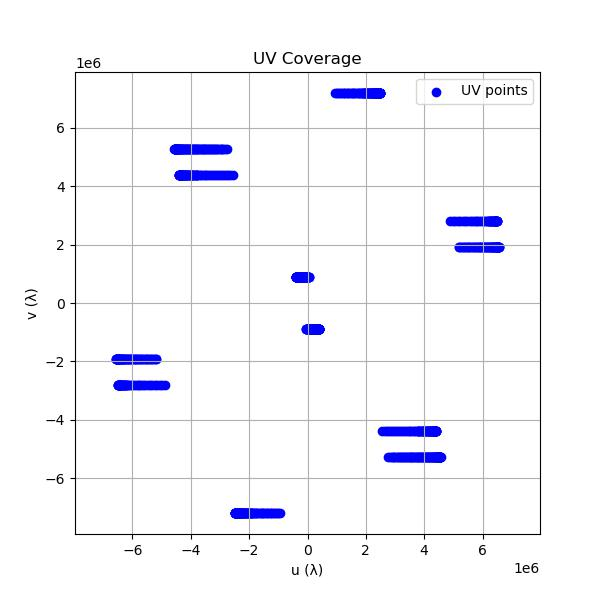}
    \caption{Dec. 0 deg.}
    \label{fig:fig2}
  \end{subfigure}
  \begin{subfigure}[b]{0.3\textwidth}
    \centering
    \includegraphics[width=\textwidth]{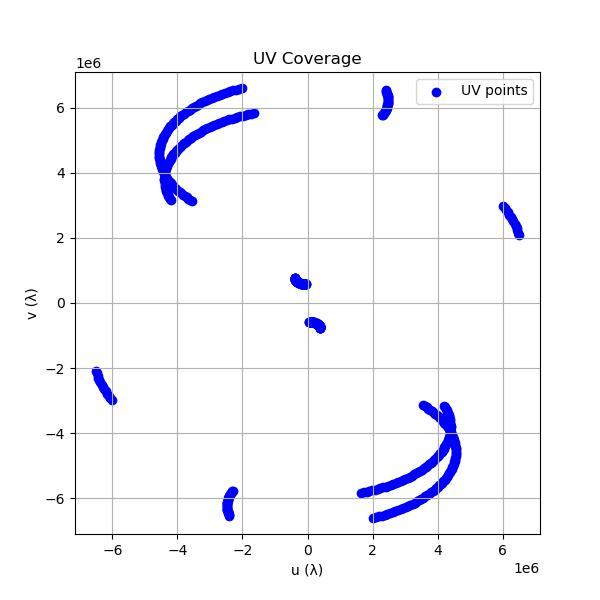}
    \caption{Dec. -30 deg.}
    \label{fig:fig3}
  \end{subfigure}
  \caption{UV coverage with SKA,GMRT,Ooty,and Iitate}
  \label{fig:three_figures}
\end{figure}

Assuming a bandwidth of 32 MHz and an integration time of 1000 seconds, the fringe detection sensitivity is estimated to be around 0.2 mJy between SKA-LOW and GMRT and 0.1 mJy between SKA-LOW and FAST at \(7\sigma\) detection. This makes it possible to conduct astrometric observations proposed in the science cases, such as foreground source detection and pulsar astrometry.

We further examined the potential for astrometry. The accuracy of astrometry is influenced by both thermal noise and systematic errors. The table 3 shows the astrometric error due to thermal noise for 1 mJy source with 5 times of 4 hours observation . At a distance of 100 pc, 1\% accuracy can be achieved with these measurements.

However, in actual observations, parallax measurements are dominated by systematic errors (\cite{kirsten2019lofar}). 

An estimation accuracy of 0.1 TEC for the ionospheric propagation path is required to achieve this precision. This may be possible through phase calibration using a single in-beam calibrator or multi-view calibration using multiple sources (\cite{rioja2020vlbi}). Previous studies using high-resolution observations with LOFAR have shown that numerous sources can be detected within the field of view (\cite{Lenc2008}), indicating a high potential for multi-view calibration.
Moreover, while the development of a robust methodology to account for the possibility that, in this frequency band, the calibrator is extended rather than point‑like remains a topic for future work, such corrections should, in principle, be implementable provided that the calibrator exhibits no significant variability over the duration of the astrometry observations.

\begin{table}[!htbp]
	\centering
	\caption{ Baseline based Astromety error (1$\sigma$ ) for 1 mJy object by thermal noise at 320MHz (mas)}
	\label{tab:astrometry_thermalnoise}
	\begin{tabular}{lcccccr} 
		\hline
		    & SKA-LOW & GMRT & Ooty & Iitate & FAST & IPS\\
		\hline
		SKA-LOW & - & 0.10 & 0.24 & 0.65 & 0.08 & 0.15 \\
		GMRT & 0.10 & -  &  2.34  & 1.08 & 0.21 & 0.35   \\
		Ooty & 0.24 & 2.34 & - &  2.22 & 0.43 & 0.68  \\
		Iitate & 0.65 & 1.08 & 2.22 & - & 1.44 & 4.93 \\
		FAST & 0.08 & 0.21 & 0.43 & 1.44 & - & 0.57 \\
		IPS & 0.15 & 0.35 & 0.68 & 4.93 & 0.57 & - \\
		\hline
	\end{tabular}
\end{table}

Furthermore, as shown in Figure 3, we consider adding antennas such as IPT and FAST, which are expected to become capable of VLBI observations in this frequency band in the near future. Figure 5 shows the UV coverage of this five-station array, which is expected to be suitable for imaging observations. The thermal noise sensitivity achievable with 8 hours of observation is estimated to be around 10 $\mu$ Jy, enabling the proposed mapping observations.

\begin{figure}[htbp]
  \centering
  \begin{subfigure}[b]{0.3\textwidth}
    \centering
    \includegraphics[width=\textwidth]{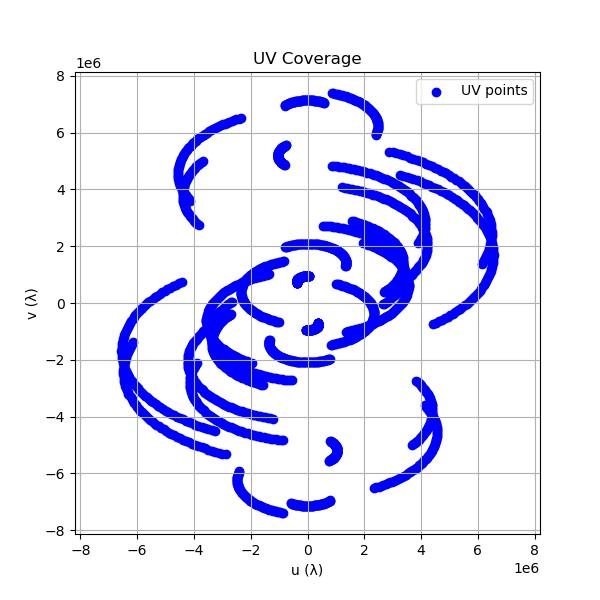}
    \caption{Dec.$+30$ deg.}
    \label{fig:fig1}
  \end{subfigure}
  \begin{subfigure}[b]{0.3\textwidth}
    \centering
    \includegraphics[width=\textwidth]{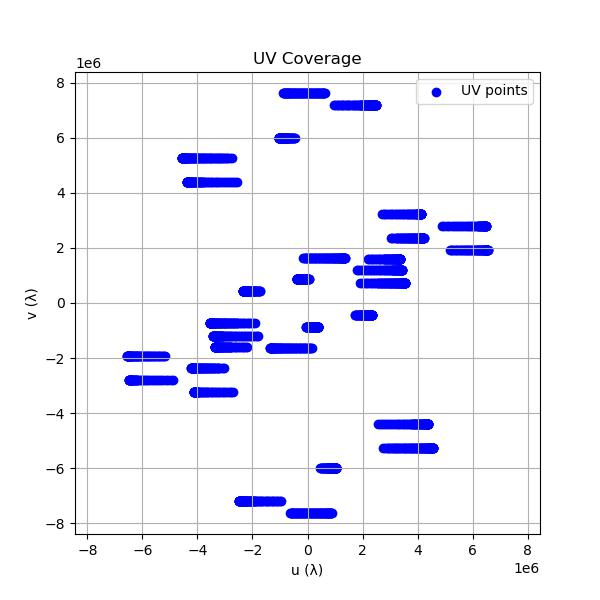}
    \caption{Dec. 0 deg.}
    \label{fig:fig2}
  \end{subfigure}
  \begin{subfigure}[b]{0.3\textwidth}
    \centering
    \includegraphics[width=\textwidth]{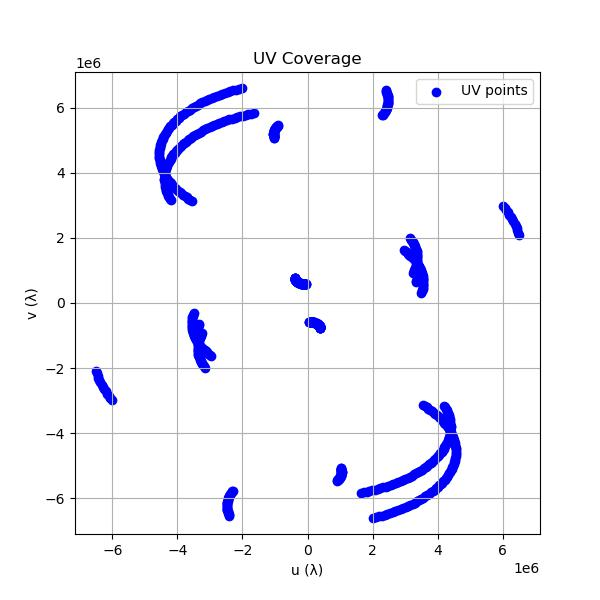}
    \caption{Dec. -30 deg.}
    \label{fig:fig3}
  \end{subfigure}
  \caption{UV coverage with SKA,GMRT,Ooty,Iitate,IPT, and FAST}
  \label{fig:three_figures}
\end{figure}

\begin{figure}[htbp]
  \centering
  \begin{subfigure}[b]{0.3\textwidth}
    \centering
    \includegraphics[width=\textwidth]{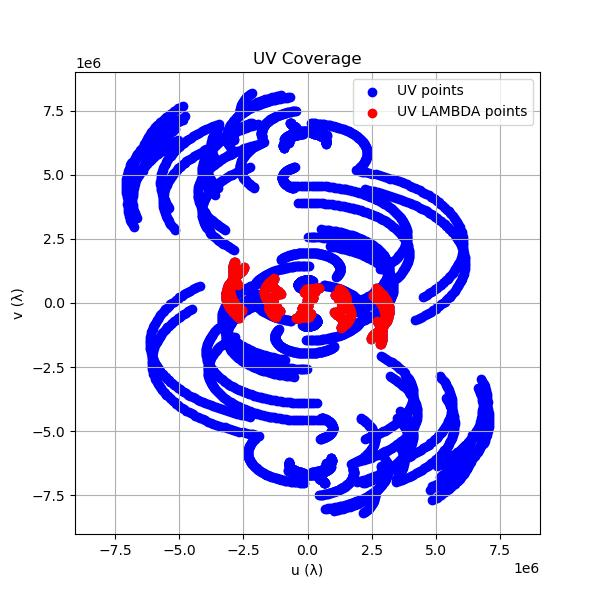}
    \caption{Dec. $+30$ deg.}
    \label{fig:fig1}
  \end{subfigure}
  \begin{subfigure}[b]{0.3\textwidth}
    \centering
    \includegraphics[width=\textwidth]{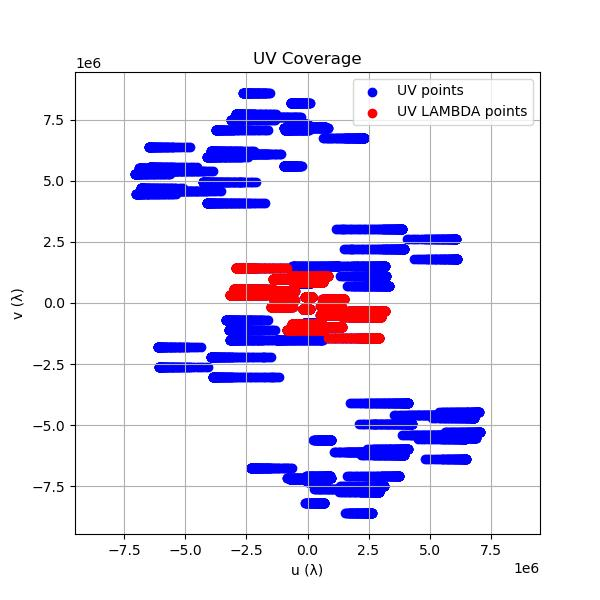}
    \caption{Dec. 0 deg.}
    \label{fig:fig2}
  \end{subfigure}
  \begin{subfigure}[b]{0.3\textwidth}
    \centering
    \includegraphics[width=\textwidth]{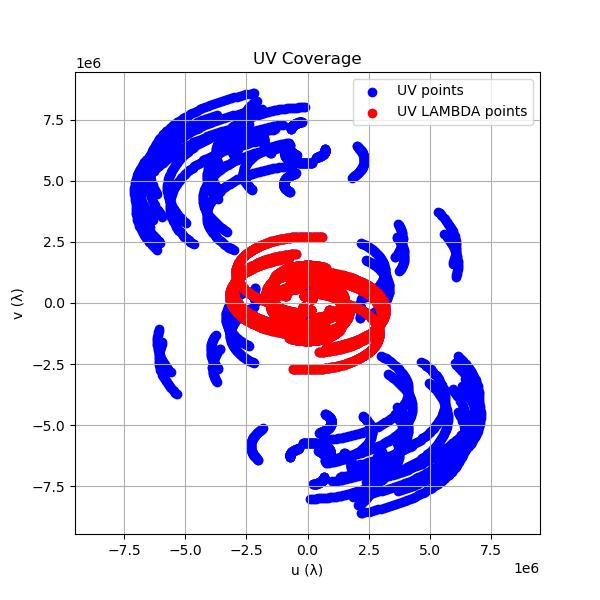}
    \caption{Dec. -30 deg.}
    \label{fig:fig3}
  \end{subfigure}
  \caption{UV coverage with SKA,GMRT,Ooty,Iitate,IPT,FAST, and LAMBDA. The red points show by SKA and LAMBDA.}
  \label{fig:three_figures}
\end{figure}

Figure 4 presents the UV coverage with ten stations including LAMBDA (Narrabri, Parkes, Ceduna, and Hobart) which is currently under development. Among them, the red points represent the UV coverage formed by SKA-LOW and LAMBDA, which are both located in Australia. This indicates that observations using SKA-LOW and LAMBDA are complementary to those that include VLBI stations outside Australia. The enhanced coverage indicates the feasibility of mapping observations with much higher dynamic range. This improvement is expected to enable high-fidelity imaging, such as detailed mapping of AGN jets and lobes, which requires superior image quality.

\begin{figure}[htbp]
    \centering
	\includegraphics[width=0.8\columnwidth]{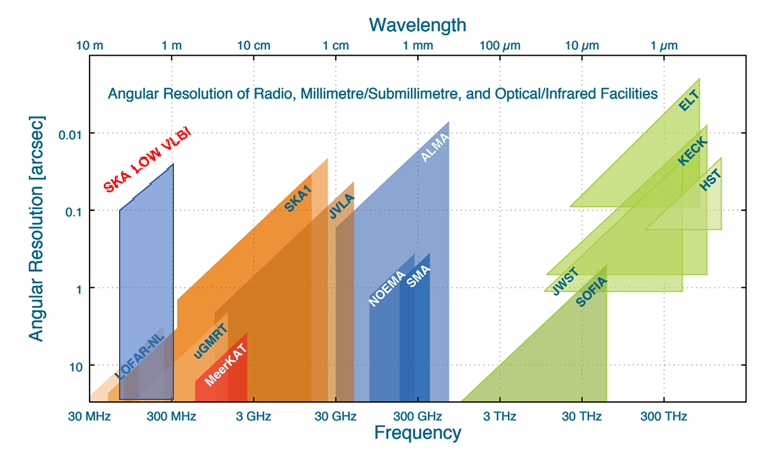}
    \caption{Spatial resolution comparison with other wavelength telescopes}
    \label{figure 5}
\end{figure}

The spatial resolution achievable with this array is comparable to that of cutting-edge telescopes at other wavelengths, such as SKA-MID, ALMA, ELT, and JWST, facilitate physical comparisons across different wavebands (see Figure 5). Conducting observations across wavelengths with consistent spatial resolution ranging from 100 to 10 milliarcseconds is of great importance in modern astronomy.

\section{Individual Antennas}
The following antennas are considered as potential elements of a low-frequency VLBI network in conjunction with SKA-LOW:

Iitate Station (Japan): Located in Iitate Village, Fukushima Prefecture, this station comprises a dual-reflector system formed by combining two parabolic dishes measuring 31 m × 16.5 m. It is currently equipped with a wideband receiver covering 150–500 MHz and a high-sensitivity receiver centered at 325 MHz (\cite{Iwai2012AMATERAS}) .

GMRT (India): The Giant Metrewave Radio Telescope, situated near Pune, consists of thirty 45-meter parabolic antennas operating as an interferometric array. It supports observations across a frequency range of 125 MHz to 1400 MHz (\cite{Gupta2017uGMRT}).

Ooty Radio Telescope (India): Constructed along a mountain slope in southern India, this instrument is a 30 m × 530 m cylindrical paraboloid. It operates at a fixed frequency of 326.5 MHz (\cite{Subrahmanya2017OWFA}).

FAST (China): The Five-hundred-meter Aperture Spherical Telescope, located in Guizhou Province, is a fixed spherical reflector with an illuminated aperture of 300 m for tracking celestial sources. While current observations are conducted using an L-band Phased Array Feed (PAF), the system is planned to support a frequency range of 70–3000 MHz (\cite{Nan2011FAST}).

IPS Array (China): Located in Inner Mongolia, this facility is designed for interplanetary scintillation observations. It consists of three 140 m × 40 m cylindrical antennas operated in phased array mode. Observations are conducted at 327, 654, and 1420 MHz (\cite{Yan2026IPSarray}). 

LAMBDA: In Australia, a long-baseline interferometric array project is underway with SKA-LOW as its foundation. The LAMDA project aims to install stations composed of 256 log-periodic antennas for SKA-Low at CSIRO’s VLBI sites, including the Parkes radio telescope, the Ceduna radio telescope, the Australia Telescope Compact Array, and the Hobart radio telescope, in order to conduct VLBI observations (\cite{reynolds2024lambda}). For this purpose, a new state-of-the-art backend system e.g., “BlueRing” and “equivalent RFSoC‑based beamformer.”, which are different from that of the SKA, is planned to be developed and implemented. The expected UV coverage, including LAMBDA, is illustrated in Figure 4. Once operational, this array is anticipated to achieve extremely high dynamic range, enabling scientific investigations that require both high spatial resolution and high dynamic range. The SEFD of each LAMBDA station was estimated from the measurement results of a SKA-Low prototype station (\cite{macario2022characterization}), and was assumed to be 2,500 Jy at 320 MHz

\section{Status of Test Observations}
Pilot VLBI observations have been initiated using the Iitate, GMRT, and Ooty stations. In 2021, a successful test observation between Iitate and Ooty resulted in the detection of clear fringes from a few objects. Further observations were conducted in May 2025, during which fringes were detected from continuum sources such as 3C147, 3C273, and others. Detailed data analysis is currently underway to evaluate system performance and validate the observational capabilities of the proposed network. Figure 6 shows fringes for each baseline of Iitate, Ooty and GMRT observation.
Observation bandwidth is 16 MHz and integration time is 59 seconds. For the source 3C147, the fringe SNRs for each baseline were 22.8, 55.4, and 1376 for Iitate–Ooty, Iitate–GMRT, and Ooty–GMRT respectively.

\begin{figure}[htbp]
    \centering
	\includegraphics[width=0.9 \columnwidth]{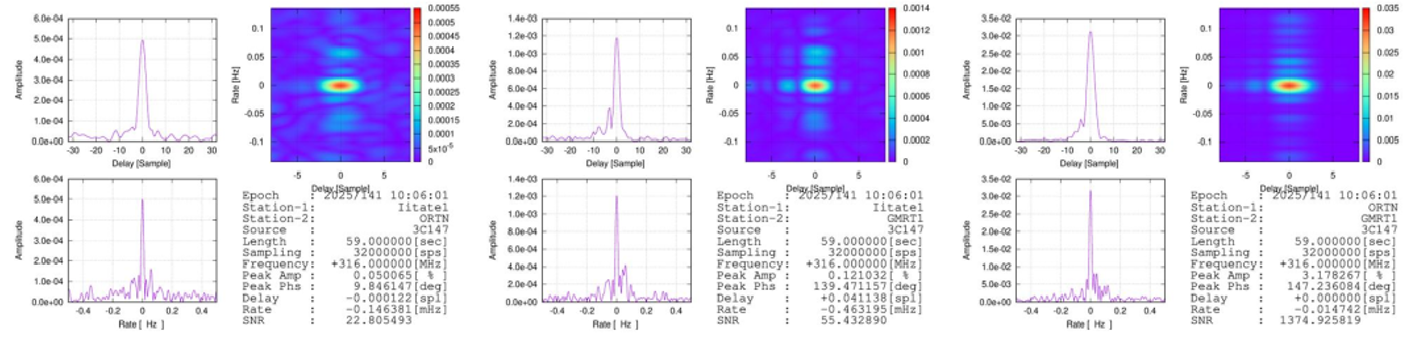}
    \caption{Fringes between Iitate, GMRT and Ooty telescopes}
    \label{figure 6}
\end{figure}

\section{Operational Infrastructure and Governance}

This low-frequency VLBI network will operate in the 16–32 MHz bandwidth available at each radio telescope outside the SKA, and even in the widest case is expected to use a bandwidth of about 128 MHz. Even when observing two polarizations simultaneously, the VLBI data recording rate will be at most around 1 Gbps, making it possible to use existing VLBI recording systems. A unified VLBI data format, VDIF (VLBI Data Interchange Format), is assumed to be employed. The data volume can also be handled with state-of-the-art software correlators such as DiFX (\cite{Deller2007DiFX}; \cite{Deller2011VLBA}). In addition, compared to other SKA observing modes, VLBI observations are initially expected to occupy only a few percent of total observing time, so it should be entirely feasible to transfer data to the current VLBI correlation centers via standard internet connections.
Regarding the organization of the network, the establishment of a Global VLBI Alliance using the SKA for VLBI is under consideration (\cite{ColomerKobayashi2019_GVA}). Within this framework, it is expected that responsibilities for VLBI operations, correlation processing, and user support will be shared.

\section{Conclusion}
We propose a VLBI network centered on SKA LOW, incorporating GMRT, Ooty, Iitate, FAST, and IPT as VLBI stations which are in the Asia-Pacific region. This network offers good capabilities for sub millli-arc-second astrometric accuracy observations, and is expected to yield significant scientific outcomes in areas such as the construction of detailed radio source catalogs for foreground subtraction in hydrogen line detection in the epoch of reionization, and parallax measurements of Pulsar timing array pulsars. Furthermore, the addition of high-sensitivity antennas and the planned LAMBDA array will enable high dynamic range observations, paving the way for new scientific investigations.

\bibliographystyle{abbrvnat-maxbibnames4}
\bibliography{reference2.bib} 


\end{document}